\documentclass[prd,preprint]{revtex4}
\usepackage{amssymb}

\begin{document}

\title{Fermionic zero-modes in type II fivebrane backgrounds}

\author{Noriaki Kitazawa}
\email{kitazawa@phys.metro-u.ac.jp}
\affiliation{Department of Physics, Tokyo Metropolitan University,
             Hachioji, Tokyo 192-0397, Japan}


\begin{abstract}

The explicit form of the fermionic zero-modes
 in the fivebrane backgrounds of type IIA and IIB supergravity theories
 is investigated.
In type IIA fivebrane background
 there are four zero-modes of gravitinos and dilatinos.
In type IIB fivebrane background
 four zero-modes of dilatinos and no zero-modes of gravitinos are found.
These zero-modes indicate the four-fermion condensates
 which have been suggested in a calculation of the tension of
 the D-brane in fivebrane backgrounds.
 
\end{abstract}

\pacs{}

\vspace*{3cm}

\maketitle

\section{Introduction}
\label{sec:intro}

The study of the non-perturbative dynamics in the superstring theory
 is still underway.
Especially,
 the dynamics of the fermionic sector is less known
 than the one in the bosonic sector.
The analysis in the low-energy effective supergravity theories
 are mainly on the bosonic sector,
 and usually the fermionic sector is not considered directly.
For example,
 although an interesting possibility of the supersymmetry breaking
 due to the gravitino condensation has been proposed
 in Refs.\cite{Witten,KMP,Rey},
 it has not been clarified yet whether this is really possible or not.
This kind of dynamics,
 which can be a non-perturbative effect of the quantum gravity,
 is very interesting in view of the low-energy physics
 because of its less model dependence.

An concrete model of the dynamical supersymmetry breaking
 due to the gravitino condensation is given in Ref.\cite{KMP}.
In the model
 the gravitino pair condensate is formed
 by the non-trivial gravitational background (Eguchi-Hanson metric),
 and the condensate triggers supersymmetry breaking
 through the Konishi anomaly relation.
Since the analysis is based on the Euclidean space-time
 and the meaning of the non-trivial gravitational background
 is unclear
 (whether a gravitational instanton or
 an Euclidean configuration in the path integral),
 it is natural to ask whether the similar dynamics
 is possible in superstring theories.

The first step was the study in the low-energy supergravity theories.
In Ref.\cite{Rey}
 a background solution of the heterotic supergravity theory,
 which is very similar to the fivebrane background in Ref.\cite{CHS1},
 is constructed, and the zero-mode solutions of gravitino and dilatino
 are found.
Although these zero-modes suggests
 the dynamical formation of the gravitino condensation,
 the condensation can not trigger the supersymmetry breaking
 through the mechanism of Ref.\cite{KMP},
 since the ten-dimensional heterotic supergravity theory
 has no global chiral symmetry.
The analysis of the fermionic zero-modes
 in fivebrane backgrounds of the heterotic supergravity theory
 has been done in Refs.\cite{Bellisai,KK}.
In Ref.\cite{KK} the author concluded that
 the gravitino condensation is possible,
 but it does not trigger the supersymmetry breaking. 

Very few analysis
 which is based on the string world-sheet theory has been achieved.
In Ref.\cite{Kitazawa}
 the author suggests some four-fermion condensations
 in fivebrane backgrounds of the type IIB string theory.
The fivebrane backgrounds
 in the type IIB supergravity theory\cite{CHS2}
 can be described by the world-sheet conformal field theory
 as a solution of the type II superstring theory
 in a certain limit\cite{AFK}.
The existence of the four fermionic zero-modes in fivebrane backgrounds
 in type IIA and type IIB supergravity theories
 is well known\cite{CHS2},
 and this number is consistent with the suggestion
 in Ref.\cite{Kitazawa}.
Although such four-fermion condensations
 do not mean the dynamical supersymmetry breaking
 through the Konishi anomaly relation
 (there are no global chiral symmetry in type IIA and IIB
 supergravity theories),
 this kind of analysis is worth doing
 to understand the non-perturbative dynamics
 in the fermionic sector of superstring theories.

In this letter
 we present the analysis based on
 the low-energy type IIA and IIB supergravity theories.
The aim of this analysis
 is to know the explicit form of four fermionic zero-modes
 in fivebrane backgrounds in type IIA and IIB supergravity theories.
The fermion field equations in fivebrane backgrounds
 in each system are explicitly solved.
In the next section
 we find four zero-modes of gravitinos and dilatinos
 in type IIA fivebrane backgrounds.
In section \ref{sec:IIB}
 four dilatino zero-modes and no gravitino zero-modes are found
 in type IIB fivebrane backgrounds.
In the last section a summary of the result is given.

\section{Fermion zero-modes in Type IIA fivebrane backgrounds}
\label{sec:IIA}

The Lagrangian of the type IIA supergravity theory
 (non-chiral $N=2$, $D=10$ supergravity theory)
 is derived from the $N=1$, $D=11$ supergravity theory
 by the dimensional reduction\cite{GP,CW,HN}.
\begin{eqnarray}
 {\cal L} = e \ e^{-2\Phi} 
&\Bigg\{&
  - {1 \over 2} R(e,\omega(e))
  + 2 \partial^\mu \Phi \partial_\mu \Phi
  - {1 \over 6} H^{\mu\nu\rho} H_{\mu\nu\rho}
  - {1 \over 2} {\bar \psi}_\mu \Gamma^{\mu\rho\nu} D_\rho \psi_\nu
  - {1 \over 2} {\bar \lambda} \Gamma^\rho D_\rho \lambda
\nonumber\\
&&
  - {\bar \psi}_\mu \Gamma^\mu \psi_\nu \partial^\nu \Phi
  + {\sqrt{2} \over 4}
    {\bar \psi}_\mu \Gamma^\nu \Gamma^\mu \Gamma_{11} \lambda
    \partial_\nu \Phi
\nonumber\\
&&
  + {1 \over {24}} H_{\alpha\beta\gamma}
    \left(
     {\bar \psi}_\mu
      \Gamma^{\mu\alpha\beta\gamma\nu} \Gamma_{11} \psi_\nu
     + 6 {\bar \psi}^\alpha \Gamma^\beta \Gamma_{11} \psi^\gamma
     - \sqrt{2} {\bar \psi}_\mu
       \Gamma^{\alpha\beta\gamma} \Gamma^\mu \lambda
    \right)
 \Bigg\}
\nonumber\\
&&
 + \left( (\mbox{fermion})^4 \ \mbox{terms}) \right),
\label{Lag-IIA}
\end{eqnarray}
 where we set all R-R fields, $C_\mu$ and $C_{\mu\nu\rho}$, zero.
We use the convention of Ref.\cite{NdRdWvN} in this section.
Here,
 $\Phi$ is the dilaton scalar field,
 $R(e,\omega(e))$ is the scalar curvature
 composed by the zhenbein $e^a{}_\mu$
 and spin connection $\omega_\mu{}^{ab}(e)$
 with the space-time coordinate $\mu$
 and local Lorentz coordinates $a$ and $b$,
 $H_{\mu\nu\rho}$ is the field strength
 of the two-form potential field $B_{\mu\nu}$,
 and $\psi_\mu$ and $\lambda$
 are the gravitino and dilatino fields, respectively,
 which are Majorana spinor including both chirality components.

The fivebrane background
 is the solution of the field equations of this Lagrangian
 with vanishing R-R fields and fermion fields.
The explicit from of the background is
\begin{eqnarray}
 H^{abc} &=& - \epsilon^{abcd} \partial_d \Phi,
\label{H-background}
\\
 e^a{}_\mu &=& e^\Phi \delta^a_\mu,
\\
 e^{2\Phi} &=& e^{2\Phi_0} + {Q \over {r^2}},
\label{Phi-background}
\end{eqnarray}
 where the local Lorentz indices $a,b,c$ and $d$ run from $6$ to $9$
 (transverse direction to the fivebrane),
 $\Phi_0$ is a constant,
 $Q$ is a constant which is quantized to be an integer,
 and $r^2 \equiv \sum_{a=6,7,8,9} x^a x^a$
 is the squared distance from the center of the fivebrane.
The dilaton background satisfies the condition
 $\square e^{2\Phi} = 0$,
 where $\square \equiv \delta_{ab} \partial_a \partial_b$
 is the four-dimensional Laplacian.
This background keeps half of the original (global) supersymmetry
 and breaks the other half of supersymmetry.
We can check this fact
 by considering infinitesimal supersymmetry variations
 with parameter $\eta(x)$.
Since all the fermion background fields vanish,
 the variation of the bosonic fields is automatically zero.
The variation of the fermionic fields are non-trivial.
\begin{eqnarray}
 \delta_\eta \lambda_{\mp}
 &=& \mp {\sqrt{2} \over 4}
     \partial_a \Phi \Gamma^a \left( 1 \mp \gamma_5 \right) \eta_{\pm},
\label{delta-dilatino-IIA}
\\
 \delta_\eta \psi_{\pm\mu}
 &=& \partial_\mu \eta_{\pm}
  + {1 \over 8} \Omega_{\mp\mu}{}^{ab} \Gamma_{ab}
    \left( 1 \mp \gamma_5 \right) \eta_{\pm}
  - {1 \over 8} \Gamma_\mu \partial_a \Phi \Gamma^a
    \left( 1 \mp \gamma_5 \right) \eta_{\pm},
\label{delta-gravitino-IIA}
\end{eqnarray}
 where
\begin{equation}
 \Gamma_{11} \eta_{\pm} = \pm \eta_{\pm}
\end{equation}
 and $\gamma_5 \equiv \Gamma^6 \Gamma^7 \Gamma^8 \Gamma^9$
 is the four-dimensional chirality operator.
The modified connection
\begin{equation}
 \Omega_{\pm\mu}{}^{ab} \equiv \omega_\mu{}^{ab} \pm H_\mu{}^{ab}
\end{equation}
 satisfies the self-dual relation:
\begin{equation}
 \Omega_{\pm\mu}{}^{ab}
  = \mp {1 \over 2} \epsilon^{abcd} \Omega_{\pm\mu cd}.
\end{equation}
From Eqs.(\ref{delta-dilatino-IIA}) and (\ref{delta-gravitino-IIA}),
 we see that
 the global supersymmetry with $\eta^{(\pm)}_\pm$ is not broken,
 but the one with $\eta^{(\mp)}_\pm$ is broken,
 where $\gamma_5 \eta^{(\pm)} = \pm \eta^{(\pm)}$.

It is easy to derive the fermion field equations
 form the Lagrangian of Eq.(\ref{Lag-IIA}).
\begin{equation}
 \Gamma^\mu \left( D_\mu - \partial_\mu \Phi \right) \lambda_\mp
 + {\sqrt{2} \over 4} \Gamma^\mu
   \left(
    \pm \partial_\nu \Phi \Gamma^\nu
    - {1 \over 6} H_{\alpha\beta\gamma} \Gamma^{\alpha\beta\gamma}
   \right) \psi_{\pm\mu} = 0,
\end{equation}
\begin{eqnarray}
 \Gamma^{\mu\rho\nu}
  \left( D_\rho - \partial_\rho \Phi \right) \psi_{\pm\nu}
&&
 + \partial^\nu \Phi \Gamma^\mu \psi_{\pm\nu}
 - \partial^\mu \Phi \Gamma^\nu \psi_{\pm\nu}
\nonumber\\
&&
 \mp {1 \over {12}}
   H_{\alpha\beta\gamma} \Gamma^{\mu\alpha\beta\gamma\nu} \psi_{\pm\nu}
 \mp {1 \over 2} H^{\mu\rho\nu} \Gamma_\rho \psi_{\pm\nu}
\nonumber\\
&&
 + {\sqrt{2} \over 4}
   \left(
    \pm \partial_\nu \Phi \Gamma^\nu
    + {1 \over 6} H_{\alpha\beta\gamma} \Gamma^{\alpha\beta\gamma}
   \right) \Gamma^\mu \lambda_{\mp} = 0.
\end{eqnarray}
In order to find the fermion zero-modes in fivebrane backgrounds,
 we only consider the chirality components of
 $\psi_{\pm\mu}^{(\mp)}$ and $\lambda_{\mp}^{(\pm)}$
 which can be induced by the broken supersymmetry variation
 with parameters $\eta^{(\mp)}_\pm$
 in Eqs.(\ref{delta-dilatino-IIA}) and (\ref{delta-gravitino-IIA}).
Furthermore,
 only the four-dimensional space of $\mu=6,7,8,9$,
 where the metric is non-trivial, is considered,
 and all the fermion fields are considered
 just as four-dimensional spinor fields.
The fermion fields are
 independent of the coordinates of
 the six-dimensional space-time $\mu=0,\cdots,5$,
 and the value of the fields vanishes
 when their indices take the values of the six-dimensional space-time.
The relevant field equations are as follows.
\begin{equation}
 \Gamma^\rho \left( D_\rho - \partial_\rho \Phi \right)
 \lambda_{\mp}^{(\pm)}
 \pm {\sqrt{2} \over 2} \partial_\rho \Phi \Gamma^\nu \Gamma^\rho
 \psi_{\pm\nu}^{(\mp)} = 0,
\label{dilatino-eq-IIA}
\end{equation}
\begin{equation}
 \left\{
 \Gamma^{\mu\rho\nu}
  \left( D_\rho - {3 \over 2} \partial_\rho \Phi \right)
  + \partial^\nu \Phi \Gamma^\mu - \partial^\mu \Phi \Gamma^\nu
 \right\}
 \psi_{\pm\nu}^{(\mp)} = 0,
\label{gravitino-eq-IIA}
\end{equation}
 where we used Eq.(\ref{H-background}).

It is naively expected that
 the supersymmetry variation
\begin{eqnarray}
 \lambda'{}_{\mp}^{(\pm)}
 &=& \mp {\sqrt{2} \over 2}
     \partial_\rho \Phi \Gamma^\rho \eta_{\pm}^{(\mp)},
\label{delta-dilatino-IIA2}
\\
 \psi'{}_{\pm\mu}
 &=& \partial_\mu \eta_{\pm}^{(\mp)}
  + {3 \over 4} \partial_\rho \Phi \Gamma_\mu{}^\rho \eta_{\pm}^{(\mp)}
  - {1 \over 4} \partial_\mu \Phi \eta_{\pm}^{(\mp)},
\label{delta-gravitino-IIA2}
\end{eqnarray}
 is the zero-mode solution.
In fact,
 it is easily shown that Eq.(\ref{dilatino-eq-IIA}) is satisfied
 for any functions $\eta_{\pm}^{(\mp)}$
 by using the relation
\begin{equation}
 \partial^\rho \partial_\rho \Phi
 + 2 \partial^\rho \Phi \partial_\rho \Phi = 0
\end{equation}
 which follows from the condition $\square e^{2\Phi} = 0$.
However, Eq.(\ref{gravitino-eq-IIA}) is not satisfied
 by Eqs.(\ref{delta-dilatino-IIA2}) and (\ref{delta-gravitino-IIA2}),
 and some modifications are required.
In case of the constant $\eta_{\pm}^{(\mp)}$,
 both Eqs.(\ref{dilatino-eq-IIA}) and (\ref{gravitino-eq-IIA})
 can be satisfied by the modification of the gravitino expression.
\begin{eqnarray}
 \lambda_0{}_{\mp}^{(\pm)}
 &=& \mp {\sqrt{2} \over 2}
     \partial_\rho \Phi \Gamma^\rho \eta_{\pm}^{(\mp)},
\label{dilatino-sol-IIA}
\\
 \psi_0{}_{\pm\mu}
 &=& \psi'{}_{\pm\mu} 
 + {3 \over 2} 
   \left(
    \partial^\rho \Phi \Gamma_{\mu\rho} \eta_{\pm}^{(\mp)}
    - {3 \over 2} \partial^\rho \Phi
                  \Gamma_\mu \Gamma_\rho \eta_{\pm}^{(\mp)}
   \right)
 = - {5 \over 2} \partial_\mu \Phi \eta_{\pm}^{(\mp)}.
\label{gravitino-sol-IIA}
\end{eqnarray}
This is the normalizable fermionic zero-mode solution
 in fivebrane background.
The gravitino solution satisfies the gauge condition
 $D^\mu \psi_0{}_{\pm\mu} = 0$.
This solution can be rewritten as follows
 by rescaling the constant parameters $\eta_{\pm}^{(\mp)}$.
\begin{eqnarray}
 \lambda_0{}_{\mp}^{(\pm)}
 &=& \mp {\sqrt{2} \over 5}
      \partial_\rho \Phi \Gamma^\rho \eta_{\pm}^{(\mp)}
  =  \lambda'{}_{\mp}^{(\pm)}
     \pm {\sqrt{2} \over 2} {3 \over 5}
     \partial_\rho \Phi \Gamma^\rho \eta_{\pm}^{(\mp)},
\label{dilatino-sol-IIA}
\\
 \psi_0{}_{\pm\mu}
 &=& - \partial_\mu \Phi \eta_{\pm}^{(\mp)}
 = \psi'{}_{\pm\mu} 
 - {3 \over 4} 
   \partial_\rho \Phi \Gamma_\mu \Gamma^\rho \eta_{\pm}^{(\mp)}.
\label{gravitino-sol-IIA}
\end{eqnarray}
This solution could be understood
 as a combination of supersymmetry variations
 and superconformal variations.

The parameter $\eta$ originally has $32$ components
 as a Majorana spinor representation
 of the space-time symmetry SO$(9,1)$.
It is decomposed in two parts $16_+$ and $16_-$
 by the ten-dimensional chirality.
They are described as four representations of
 SO$(5,1)\times$SO$(4)\subset$SO$(9,1)$ as follows.
\begin{eqnarray}
 16_+ &=& (4_+,2_+) + (4_-,2_-),
\label{16+}
\\
 16_- &=& (4_+,2_-) + (4_-,2_+).
\label{16-}
\end{eqnarray}
The parameters $\eta_{+}^{(-)}$ and $\eta_{-}^{(+)}$
 correspond to the representations of
 $(4_-,2_-)$ and $(4_-,2_+)$, respectively,
 and each has two independent component as a SO$(4)$ Weyl spinor.
Therefore,
 we find the explicit form fo the
 four fermionic zero-modes in type IIA fivebrane backgrounds.

\section{Fermion zero-modes in Type IIB fivebrane backgrounds}
\label{sec:IIB}

Although there is no manifestly Lorentz-invariant action
 of the type IIB supergravity theory
 (chiral $N=2$, $D=10$ supergravity theory),
 the covariant field equations have been obtained\cite{Schwarz}.
The field equations of fermion fields are
\begin{eqnarray}
 \gamma^\mu D_\mu \lambda
 &=& {1 \over {240}} i
     \gamma^{\rho_1 \cdots \rho_5} \lambda F_{\rho_1 \cdots \rho_5}
 + \left( (\mbox{fermion})^3 \ \mbox{terms}) \right),
\label{dilatino-eq-IIB-org}
\\
 \gamma^{\mu\rho\nu} D_\rho \psi_\nu
 &=& - {1 \over 2} i \gamma^\rho \gamma^\mu \lambda^* P_\rho
     - {1 \over {48}} i \gamma^{\nu\rho\sigma} \gamma^\mu \lambda
       G^*_{\nu\rho\sigma}
 + \left( (\mbox{fermion})^3 \ \mbox{terms}) \right),
\label{gravitino-eq-IIB-org}
\end{eqnarray}
 where the convention of Ref.\cite{Schwarz} is used
 ($\kappa=1$).
Here,
 the dilatino field $\lambda = \lambda_R + i \lambda_I$
 is the complex Weyl spinor with positive ten-dimensional chirality,
 the gravitino field $\psi_\mu = \psi_{R\mu} + i \psi_{I\mu}$
 is the complex Weyl spinor with negative ten-dimensional chirality,
 $F_{\rho_1 \cdots \rho_5}$ is the field strength
 of the real four-form potential field $A_{\rho_1 \cdots \rho_4}$,
 $P_\rho$ is the ``field strength'' of the complex scalar field $B$,
 and $G_{\nu\rho\sigma}$ is the field strength
 of the complex two-form field $A_{\rho\sigma}$.
The covariant derivative includes an unusual contribution:
\begin{eqnarray}
 D_\mu \lambda
 &=& {\tilde D}_\mu \lambda - {3 \over 2} i Q_\mu \lambda,
\\
 D_\rho \psi_\nu
 &=& {\tilde D}_\rho \psi_\nu - {1 \over 2} i Q_\rho \psi_\nu,
\end{eqnarray}
 where ${\tilde D}$ denotes the usual covariant derivative
 in curved space-time and
\begin{equation}
 Q_\mu
 = {1 \over {1 - B^* B}}
   \mbox{Im} \left( B \partial_\mu B^* \right).
   \end{equation}
The supersymmetry transformation rule of these fermion fields is
\begin{eqnarray}
 \delta \lambda
 &=& i \gamma^\mu \epsilon^* {\hat P}_\mu
 - {1 \over {24}} i \gamma^{\mu\rho\sigma} \epsilon 
   {\hat G}_{\mu\rho\sigma}
 + {3 \over 2} i
   \mbox{Im} \left( B {\bar \epsilon} \lambda^* \right) \lambda,
\\
 \delta \psi_\mu
 &=& D_\mu \epsilon
 + {1 \over {480}} i \gamma^{\rho_1 \cdots \rho_5} \gamma_\mu \epsilon
   {\hat F}_{\rho_1 \cdots \rho_5}
 + {1 \over {96}}
   \left(
    \gamma_\mu{}^{\nu\rho\sigma} {\hat G}_{\nu\rho\sigma}
    - 9 \gamma^{\rho\sigma} {\hat G}_{\mu\rho\sigma}
   \right) \epsilon^*
\nonumber\\
&&
 + \left( (\mbox{fermion})^2 \ \mbox{terms}) \right),
\end{eqnarray}
 where the parameter $\epsilon = \epsilon_R + i \epsilon_I$
 is the complex Weyl spinor with negative ten-dimensional chirality
 with
\begin{equation}
 D_\mu \epsilon
 = {\tilde D}_\mu \epsilon - {1 \over 2} i Q_\mu \epsilon,
\end{equation}
 and hat means the supercovariantization.

The fivebrane background is a classical configuration
 which preserves half of the original (global) supersymmetry.
The background field configuration is as follows\cite{CHS2}.
For $P_\mu$, $G_{\nu\rho\sigma}$ and $F_{\rho_1 \cdots \rho_5}$ fields,
\begin{eqnarray}
 P_\mu &=& {1 \over 2} \partial_\mu \Phi,
\label{P-background}
\\
 G_{abc} &=& - 2 \epsilon_{abcd} \partial^d \Phi,
\\
 F_{\rho_1 \cdots \rho_5} &=& 0,
\end{eqnarray}
 where $a,b,c$ and $d$
 are the local Lorentz coordinates which run from $6$ to $9$
 and $\Phi$ is the real scalar field
 satisfying Eq.(\ref{Phi-background}).
Eq.(\ref{P-background})
 can be understood as the background $B = B_R + i B_I$ field of
\begin{eqnarray}
 B_R &=& \tanh \Phi/2,
\\
 B_I &=& 0,
\end{eqnarray}
 and this results $Q_\mu = 0$.
The background metric is
\begin{equation}
 e^a{}_\mu =
 \left\{
 \begin{array}{cc}
  e^{-\Phi/4} \delta^a_\mu, & \qquad a,\mu = 0, \cdots 5, \\
  e^{3\Phi/4} \delta^a_\mu, & \qquad a,\mu = 6, \cdots 9, \\
  0, & \qquad \mbox{other} \ a \ \mbox{and} \ \mu.
 \end{array}
 \right.
\end{equation}
Note that we are using the Einstein metric.

The supersymmetry variation of bosonic fields vanish,
 since all the fermionic background fields are zero.
The supersymmetry variation of fermionic fields is non-trivial.
\begin{eqnarray}
 \delta \lambda
 &=& \partial_\mu \Phi i \gamma^\mu
     {{1-\Gamma_5} \over 2} \epsilon_R
 - i \partial_\mu \Phi i \gamma^\mu
     {{1+\Gamma_5} \over 2} \epsilon_I,
\label{delta-dilatino-IIB}
\\
 \delta \psi_\mu
 &=& \left(
      \partial_\mu + {1 \over 8} \partial_\mu \Phi \Gamma_5
     \right) \epsilon_R
 + i \left(
      \partial_\mu - {1 \over 8} \partial_\mu \Phi \Gamma_5
     \right) \epsilon_I
\nonumber\\
&&
   + {3 \over 4} \partial^\nu \Phi \gamma_{\mu\nu}
     {{1-\Gamma_5} \over 2} \epsilon_R
 + i {3 \over 4} \partial^\nu \Phi \gamma_{\mu\nu}
     {{1+\Gamma_5} \over 2} \epsilon_I,
\label{delta-gravitino-IIB}
\end{eqnarray}
 where $\Gamma_5 \equiv \gamma^6 \gamma^7 \gamma^8 \gamma^9$
 is the four-dimensional chirality operator.
Note that all $\gamma^\mu$ are imaginary.
We see that the supersymmetry with
 $\epsilon_R^{(-)}$ and $\epsilon_I^{(+)}$ is broken,
 and the supersymmetry with
 $\epsilon_R^{(+)} = e^{-\Phi/8} \eta_R^{(+)}$
 and $\epsilon_I^{(-)} = e^{-\Phi/8} \eta_I^{(-)}$ is preserved,
 where $\Gamma_5 \epsilon_{R,I}^{(\pm)} = \pm \epsilon_{R,I}^{(\pm)}$
 and $\eta_R^{(+)}$ and $\eta_I^{(-)}$ are constant spinors.

The fermion field equations in fivebrane backgrounds are simple.
From Eqs.(\ref{dilatino-eq-IIB-org}) and (\ref{gravitino-eq-IIB-org}),
\begin{eqnarray}
 \gamma^\mu D_\mu \lambda &=& 0,
\\
 \gamma^{\mu\rho\nu} D_\rho \psi_\nu
 &+& {1 \over 4} i \partial_\rho \Phi \gamma^\rho \gamma^\mu
   \left( \lambda^* - \Gamma_5 \lambda \right) = 0.
\end{eqnarray}
In order to find the fermionic zero-modes in fivebrane backgrounds,
 we only consider the fermion components
 $\psi_{R\mu}^{(-)}$, $\psi_{I\mu}^{(+)}$,
 $\lambda_R^{(+)}$ and $\lambda_I^{(-)}$
 which can be induced by the broken supersymmetry variation.
Furthermore, we take the following ansatz:
 only the four-dimensional space of $\mu=6,7,8,9$
 should be considered,
 and all the fermion fields are considered
 just as four-dimensional spinor fields.
The fermion fields are
 independent of the coordinates of
 the six-dimensional space-time $\mu=0,\cdots,5$,
 and the value of fields vanishes
 when their indices take the values of the six-dimensional space-time.
Then, the field equations become quite simple.
\begin{eqnarray}
 \gamma^\mu D_\mu \lambda
 &=& \gamma^\mu
     \left(
      \partial_\mu + {9 \over 8} \partial_\mu \Phi
     \right) \lambda = 0,
\label{dilatino-eq-IIB}
\\
 \gamma^{\mu\rho\nu} D_\rho \psi_\nu
 &=& \gamma^{\mu\rho\nu}
     \left(
      \partial_\mu
      + {3 \over 8} \partial^\sigma \Phi \gamma_{\rho\sigma}
     \right) \psi_\nu = 0,
\label{gravitino-eq-IIB}
\end{eqnarray}
 where $\lambda$ can be $\lambda_R^{(+)}$ or $\lambda_I^{(-)}$
 and $\psi_\nu$ can be $\psi_{R\nu}^{(-)}$ or $\psi_{I\nu}^{(+)}$.

The candidate of the solution of
 Eqs.(\ref{dilatino-eq-IIB}) and (\ref{gravitino-eq-IIB})
 is the following broken supersymmetry variation.
\begin{eqnarray}
 \lambda'
 &=& \partial_\mu \Phi i \gamma^\mu \epsilon_R^{(-)}
 - i \partial_\mu \Phi i \gamma^\mu \epsilon_I^{(+)},
\label{delta-dilatino-IIB2}
\\
 \psi'{}_\mu
 &=& \left(
      \partial_\mu - {1 \over 8} \partial_\mu \Phi
      + {3 \over 4} \partial^\nu \Phi \gamma_{\mu\nu}
     \right) \epsilon_R^{(-)}
 + i \left(
      \partial_\mu - {1 \over 8} \partial_\mu \Phi
      + {3 \over 4} \partial^\nu \Phi \gamma_{\mu\nu}
     \right) \epsilon_I^{(+)}.
\label{delta-gravitino-IIB2}
\end{eqnarray}
Eq.(\ref{delta-dilatino-IIB2})
 is the solution of Eq.(\ref{dilatino-eq-IIB}),
 if $\epsilon_R^{(-)} = e^{{13 \over 8} \Phi} \eta_R^{(-)}$
 and $\epsilon_I^{(+)} = e^{{13 \over 8} \Phi} \eta_I^{(+)}$,
 where $\eta_R^{(-)}$ and $\eta_I^{(+)}$ are constant spinors.
Namely, the solution is
\begin{equation}
 \lambda_0
 = \partial_\mu \Phi i \gamma^\mu e^{{13 \over 8} \Phi} \eta_R^{(-)}
 - i \partial_\mu \Phi i \gamma^\mu e^{{13 \over 8} \Phi} \eta_I^{(+)},
\end{equation}
 and this is a normalizable solution.
Eq.(\ref{delta-gravitino-IIB2})
 is not the solution of Eq.(\ref{gravitino-eq-IIB}).
The real part of Eq.(\ref{delta-gravitino-IIB2})
 is a linear combinations of all three possible independent terms
 in the lowest order of the derivative expansion.
The solution should be a linear combination of these three terms.
The same is true for the imaginary part of
 Eq.(\ref{delta-gravitino-IIB2}).
However,
 we can explicitly show that
 any linear combination can not be a solution of
 Eq.(\ref{gravitino-eq-IIB}).
Therefore,
 we conclude that there is no gravitino zero-mode
 in the lowest order of the derivative expansion.

The parameter of the supersymmetry transformation $\epsilon$
 originally has $32$ components.
Each of real and imaginary part of $\epsilon$
 belongs to the $16_-$ Majorana-Weyl representation
 of the space-time symmetry SO$(9,1)$.
Each of them is described as two representations of
 SO$(5,1)\times$SO$(4) \subset$SO$(9,1)$ as in Eq.(\ref{16-}).
The spinors $\eta_R^{(-)}$ and $\eta_I^{(+)}$
 correspond to the representations of
 $(4_+,2_-)$ and $(4_-,2_+)$, respectively,
 and each has two independent component as a SO$(4)$ spinor.
Therefore,
 we find the explicit form of the
 four dilatino zero-modes in type IIB fivebrane backgrounds.

\section{Conclusion}
\label{sec:conclusion}

We have found the explicit form of the four fermionic zero-modes
 in both type IIA and IIB fivebrane backgrounds.
In type IIA fivebrane backgrounds
 both dilatinos and gravitinos have zero-modes.
On the other hand,
 in type IIB fivebrane backgrounds
 only dilatinos have zero-modes
 and there are no gravitino zero-modes.
If we believe the naive consideration
 based on the path integral in the low-energy supergravity theory,
 this result suggests four-fermion condensations
 in fivebrane backgrounds.

In type IIB fivebrane backgrounds
 four-fermion condensations have been suggested in Ref.\cite{Kitazawa}
 through the calculation of the D3-brane tension
 in type IIB fivebrane backgrounds
 using the world-sheet conformal field theory.
To understand
 which four-fermion condensations forms,
 we need further knowledge of the higher-order terms
 in the low-energy supergravity theory.

The naive consideration based on the path integral
 also suggests that there should be no fermion pair condensates
 in fivebrane background,
 since there are four fermion zero-modes.
This can be checked
 by calculating the fermion propagator in the string theory
 using the world-sheet conformal field theory.
If this naive consideration is correct,
 there should be no massless fermion propagation.
The result of this calculation will be given in future works.

\acknowledgments

The author would like to thank S.~Saito and Y.~Katagiri
 for useful discussions.

\end{document}